\documentclass[doublespacing]{elsart}

\usepackage{amssymb}
\usepackage{graphics}
\usepackage{epsfig}

\def\onlinecite{\cite}

\begin{document}




\pagestyle{plain}
\setcounter{page}{1}
\date{\today}

\begin{frontmatter}

\title
{Selection of dominant multi-exciton transitions in disordered linear
J-aggregates}

\author{
J. A. Klugkist, V. A. Malyshev, J. Knoester$^*$}

\address
{Centre for Theoretical Physics and  Zernike Institute for Advanced
Materials,\\
University of Groningen, Nijenborgh 4, 9747 AG Groningen, The
Netherlands}

\corauth[corauthor]{ Corresponding author: J. Knoester, Centre for
Theoretical Physics and Zernike Institute for Advanced Materials,
University of Groningen, Nijenborgh 4, 9747 AG Groningen, The
Netherlands; tel: +31-50-3634369; fax: +31-50-3634947; email:
j.knoester@rug.nl }

\begin{abstract}
We show that the third-order optical response of disordered linear
J-aggregates can be calculated by considering only a limited number
of transitions between (multi-) exciton states. We calculate the
pump-probe absorption spectrum resulting from the truncated set of
transitions and show that, apart from the blue wing of the induced
absorption peak, it agrees well with the exact spectrum.
\end{abstract}


\journal{J. Luminescence}
\end{frontmatter}

\newpage

\section{Introduction}    \label{Sec: Introduction}

Since the discovery of the J-band of pseudoisocyanine
(PIC)~\cite{Jelley1937,Scheibe1937}, the  study of the collective
optical properties of molecular aggregates has received much
attention. The collective nature of the excitations gives rise to
narrow absorption lines (exchange narrowing) and ultra-fast
spontaneous emission. The development of novel optical techniques, a
continuously increasing theoretical understanding of the optical
response of these systems, and the realization that this type of
excitations play an important role in natural light harvesting
systems have kept these materials in the spotlight.

During the past 15 years a topic of particular interest has been the
effect of multi-exciton states on nonlinear optical spectroscopies.
Amongst the latter are the nonlinear absorption spectrum
\cite{Spano1989,Spano1991}, photon echoes
\cite{Spano1992,Burgel1995}, pump-probe spectroscopy
\cite{Burgel1995,Fidder1993,Bakalis1999}, and two-dimensional
spectroscopy \cite{Brixner2006,Heijs2007}. From a theoretical point
of view, accounting for multi-exciton states requires the handling
of large matrices, due to the extent of the associated Hilbert
space. It is well-known, however, that for homogeneous linear
aggregates, three states dominate the third-order response, namely
the ground state, the lowest one-exciton state and the lowest
two-exciton state. In practice, J-aggregates suffer from appreciable
disorder, which leads to localization of the exciton states. It was
shown in Ref.~\cite{Fidder1993} that the three-state picture still
holds to a good approximation, provided that one replaces the chain
length by the typical exciton localization length. The existence of
the so-called hidden structure of the Lifshits tail of the density
of states (DOS) justifies this
approach~\cite{Malyshev1995,Malyshev2001}.

In this paper, we put a firmer basis under the above idea, by
systematically analyzing the dominant ground-state to one-exciton
transitions and one- to two-exciton transitions in disordered linear
aggregates. By comparison to exact spectra, we show that, indeed,
the picture of three dominant states per localization segment holds.

\section{Model}    \label{Sec: Model}

We model a single aggregate as a linear chain of $N$ coupled  two-level
monomers with parallel transition dipoles. We assume that the aggregate
interacts with a disordered environment, resulting in random fluctuations
in the molecular
transition energies $\varepsilon_n$ (diagonal disorder), and restrict
 ourselves to nearest neighbor excitation transfer interactions $-J$.
 The optical excitations are described by the Frenkel exciton Hamiltonian,
\begin{equation}
    H = \sum_{n=1}^N  \varepsilon_n |n\rangle \langle n| -
    J\sum_{n=1}^{N-1}\left( |n\rangle \langle n+1| + h.c.\right).
\label{H}
\end{equation}
Here, $|n \rangle$ denotes the state with the $n$th site excited and
all other sites in the ground state. The monomer excitation energies
$\varepsilon_n$ are modeled as uncorrelated Gaussian variables with
zero mean and standard deviation $\sigma$. For J-aggregates $J > 0$.
Numerical diagonalization of the Hamilton yields the exciton
energies $\varepsilon_\nu$ ($\nu = 1,\ldots , N$) and exciton
wavefunctions $ |\nu\rangle = \sum_{n=1}^N \varphi_{\nu n}|n\rangle$
of the one-exciton states, where $\varphi_{\nu n}$ is the $n$th
component of the wavefunction $|\nu\rangle$. The two exciton states
are given by the Slater determinant of two different one-exciton
states $|\nu_1\rangle$ and $|\nu_2\rangle$~\cite{Spano1991}: $|
\nu_1,\nu_2 \rangle = \sum_{n\leq m}^N (\phi_{\nu_1n}\phi_{\nu_2 m}-
\phi_{\nu_1 m}\phi_{\nu_2n})| n,m\rangle$ with $|n,m\rangle$ the
state in which the sites $n$ and $m$  are excited and all other
sites are in the ground state. The corresponding two-exciton energy
is given by $E_{\nu_1\nu_2} = E_{\nu_1}+E_{\nu_2}$.

\section{Selecting dominant transitions}

\begin{figure}[lh]
\begin{center}
        \includegraphics[width = \textwidth,scale=1]{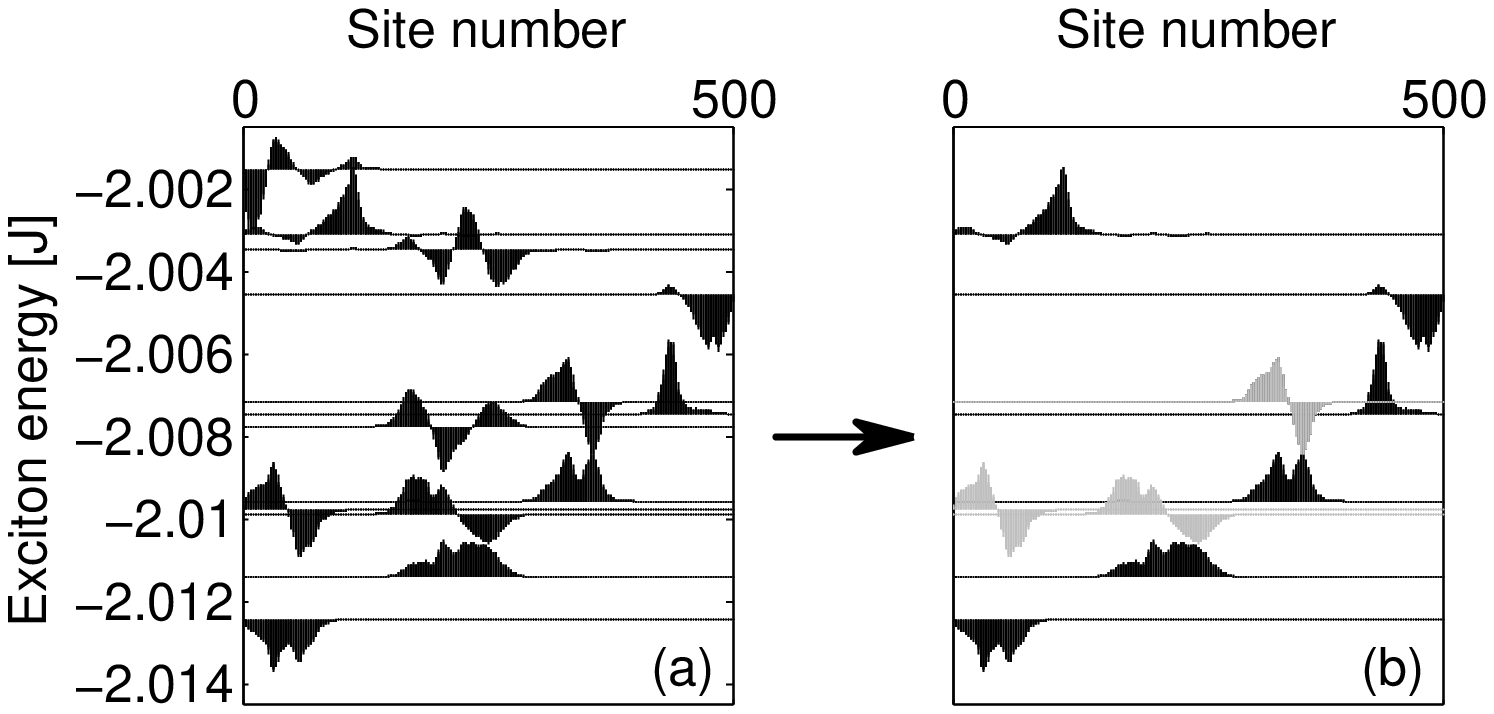}
\end{center}\caption{}\label{Fig: Reduced wavefunctions}
\caption{(a)~The lowest 12 one-exciton states of a chain of length
$N=500$ for a particular disorder realization. (b)~A subset of $s$
states (black) and $p$ states (gray) that mostly contribute to the
one-to-two exciton transitions.  }
\end{figure}

The physical size of a molecular aggregate can amount to thousands of monomers,
but the disorder localizes the exciton states~\cite{Abrahams1979}. For
J-aggregates a small number of states at the bottom of the one-exciton band
contain almost all the oscillator strength. The states are localized on
segments with a typical extension $N^*$, called the localization length,
which depends on the magnitude of the disorder~\cite{Fidder1991}. The
wavefunctions of these states overlap weakly and consist of (mainly) a
single peak (they have no node within the localization segment), see
Fig.~\ref{Fig: Reduced wavefunctions}. For the remainder of this paper we
will refer to these states as $s$ states, and $\mathcal{S}$ will denote
the set of $s$ states.

\begin{figure}[lh]
\begin{center}
        \includegraphics[width = \textwidth,scale=1]{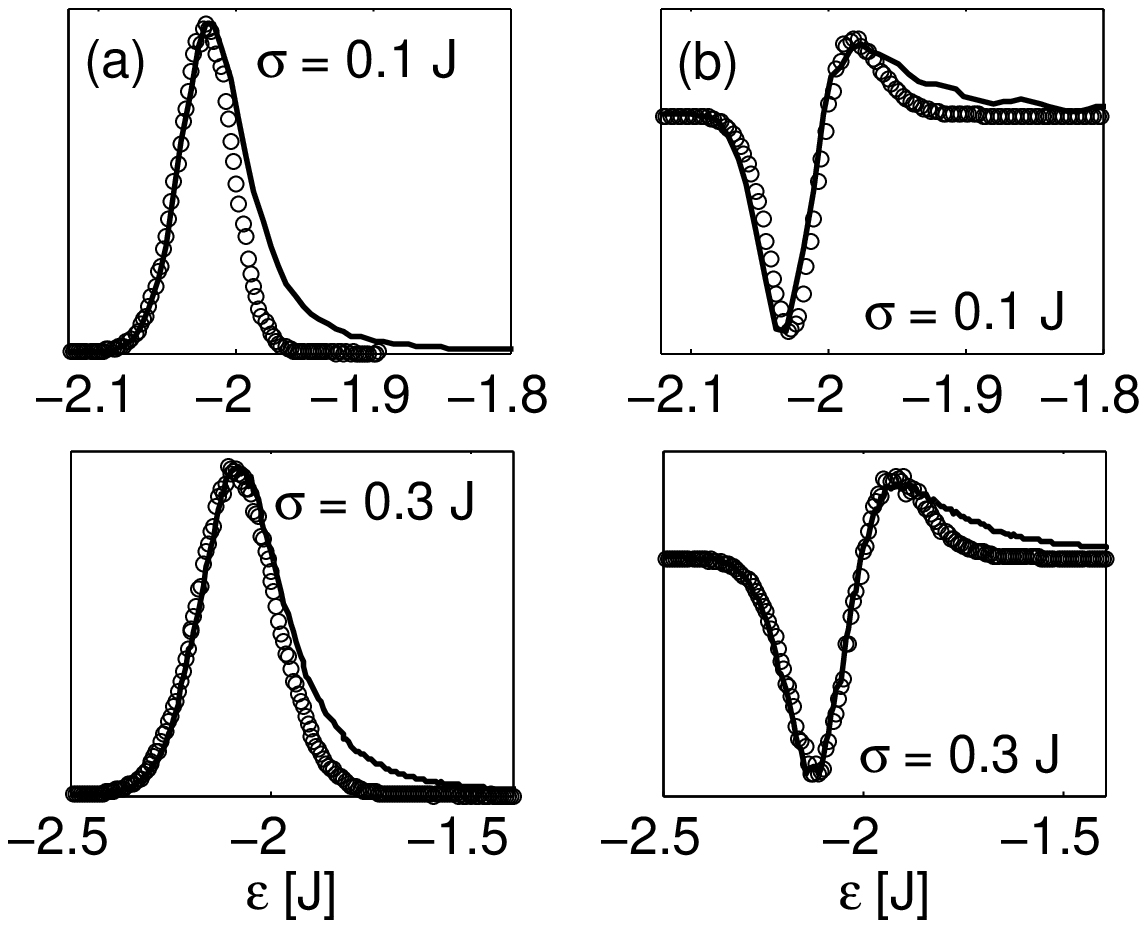}
\end{center}\caption{}\label{Fig: Absorption}
\caption{ (a)~Absorption spectrum due to $s$ states (circles)
compared to the exact one (solid line). (b)~Pump-probe spectrum due
to the selected $s$ and $p$ type states (circles) compared to the
exact one (solid line). The spectra were calculated for $\sigma=0.1
J$ (top) and $\sigma=0.3 J$ (bottom).}
\end{figure}

From the complete set of wavefunctions we  select the $s$ states
using the selection rule proposed in Ref.~\onlinecite{Malyshev2001},
$\big|\sum_n \varphi_{\nu n} |\varphi_{\nu n}|\big|  \ge C_0$. For a
disorder-free the lowest state  contains 81\% of the total
oscillator strength between the ground state and the one-exciton
band~\cite{Fidder1991}. Numerically we found that the $s$ states
selected by taking $C_0=0.75$ (for $0.05 J < \sigma<0.3 J$) together
contain 76\% of the total oscillator strength: ${\langle \sum_{s\in
\mathcal{S}}\mu_{s}^2\rangle}/{N} \approx 0.76$. Here $\mu_{s}$ is
the transition dipole moment from the ground state to the state
$|s\rangle$. In Fig.~\ref{Fig: Absorption}(a) we show the absorption
spectrum calculated only with the $s$ states and compare it  to the
exact spectrum.

We observe that the $s$ states give a good representation of this
spectrum, except for its blue wing, where higher-energy exciton
states, which contain one node within the localization segment,
contribute as well~\cite{Malyshev2007,Klugkist2007}.  These
so-called $p$ states may be identified as the second one-exciton
state on a localization segment~\cite{Malyshev1995,Malyshev2001}.

The $p$ states play a crucial role in the third-order response. In
order to analyze this, we have considered two-exciton states
$|s,p\rangle$ given by the Slater determinant of a given $s$ state
(selected as described above) with all other one-exciton states $\nu
\notin \mathcal{S}$  (the two-exciton state consisting of two $s$
type states localized on different segments do not contribute to the
nonlinear response), and calculate the corresponding transition
dipole moments $\mu_{s\nu,s}$. From the whole set of
$\mu_{s\nu,s}$, we select the largest one, denoted by
$\mu_{sp_s,s}$, were the  substrict $s$ in $p_s$ indicates its
relation with the state $|s\rangle$.  It turns out that the
one-exciton state  $|p_s\rangle$ selected in this way is localized
on the same segment as the state $|s\rangle$. Several of these
doublets of $s$ and $p$ states are shown in Fig.~\ref{Fig: Reduced
wavefunctions}(b). The partners of the lowest $s$ states indeed look
like $p$ states, having a well-defined node within the localization
segment. They form the hidden structure of the Lifshits tail of the
DOS~\cite{Malyshev1995} we mentioned above. For higher lying $s$
states, these partners (not shown) are more delocalized and often do
not have a $p$-like shape.

The average ratio of the oscillator strength of the transitions
$|0\rangle\to |s\rangle$ and $|s\rangle\to |p_s\rangle$ turned out
to be $\left\langle {\mu_{sp_s,s}^2}/{\mu_{s}^2}\right\rangle
\approx 1.4$. For the dominant ground-to-one and one-to-two exciton
transition in a homogeneous chain, this ratio reads
${\mu_{12,1}^2}/{\mu_{1}^2} \approx 1.57$. This comparison suggests
that our selection of two-exciton states well captures the dominant
one-to-two exciton transitions in disordered chains. We have also
found that the energy separation between the states $|s\rangle$ and
$|p_s\rangle$ obeys $
    \left\langle E_{s}-{E_{p_s}}\right\rangle \approx 1.35 \times{ 3\pi^2J}/{N^{*2}}
$, with $N^*  =  \left\langle{\mu_s^2}/{0.81}\right\rangle$. This
resembles the level spacing of  a homogeneous chain: $E_{2} -{E_1}
\approx{3\pi^2J}/{N^2}$, confirming that the separation between the
one-exciton bleaching peak and the one-to-two-exciton induced
absorption peak may be used to extract the typical localization size
from experiment~\cite{Bakalis1999}.

To illustrate how the selected transitions, involving the ($s$,$p$)
doublets, reproduce the optical response of the aggregate, we have
calculated the pump-probe spectrum at zero temperature using only
these transitions \cite{Heijs2007},\begin{equation}\label{Eq: PP}
P(\omega)=
\bigg\langle-2\mu_{s}^4\delta(\omega_{s}-\omega)+
\mu_{s}^2\mu_{p_s}^2\delta(\omega_{p_s}-\omega)\bigg\rangle,
\end{equation}
and compared the result to the exact spectrum, see  Fig.~\ref{Fig:
Absorption}(b). Apart from a blue wing in the induced absorption
part of the spectrum (the positive peak), the selected transitions
reproduce the exact pump-probe spectrum very well. In particular the
separation between the bleaching and induced absorption peaks is
practically identical to the exact one.

\section{Conclusion}

We have shown that the third-order optical response of disordered
linear J-aggregates is dominated by a very limited number of
transitions. Considering only these transitions enormously reduces
the computational effort necessary to simulate nonlinear
experiments. We have used the procedure outlined above to calculate
the optical bistable response of a thin film of J-aggregates, taking
into account the one-to-two exciton transitions \cite{Klugkist2007}.










\begin{thebibliography}{99}


\bibitem{Jelley1937} E. E. Jelley,  Nature {\bf 139}, 631  (1937).

\bibitem{Scheibe1937} G. Scheibe, Angew. Chem.  {\bf 50}, 212  (1937).



\bibitem{Spano1989} F. C. Spano , and S. Mukamel, Phys. Rev. A {\bf 40},
5783 (1989).



\bibitem{Spano1991} F. C. Spano, Phys. Rev. Lett. {\bf 67}, 3424 (1991).




\bibitem{Spano1992} F. C. Spano, J. Chem. Phys. {\bf 96}, 2845 (1992).




\bibitem{Burgel1995} M. van Burgel, D. A. Wiersma, and K. Duppen, J.
Chem. Phys.{\bf 102}, 20 (1995).



\bibitem{Fidder1993} H.~Fidder, J.~Knoester, and D.~A.~Wiersma,
    J. Chem. Phys. {\bf 98}, 6564 (1993).


\bibitem{Bakalis1999} L.~Bakalis,  and J. Knoester
    J. Phys. Chem. B {\bf 103}, 6620 (1999).



\bibitem{Brixner2006} T.~Brixner \textit{et al}, J. Phys. Chem. B
{\bf 110}, 20032 (2006).

\bibitem{Heijs2007} D. J. Heijs, A. G. Dijkstra, and J. Knoester,
accepted by Chem. Phys.

\bibitem{Malyshev1995} V. Malyshev and P. Moreno, Phys. Rev. B
    {\bf 51} 14587 (1995).

    \bibitem{Malyshev2001} A. V. Malyshev and V. A. Malyshev, Phys. Rev. B
    {\bf 63} 195111 (2001).











\bibitem{Abrahams1979} E. Abrahams \textit{et. al.}, Phys. Rev. Lett.
{\bf 42}, 679 (1979).

\bibitem{Fidder1991} H.~Fidder, J.~Knoester, and D.~A.~Wiersma,
    J. Chem. Phys. {\bf 95}, 7880 (1991).


\bibitem{Malyshev2007}  A. V. Malyshev, V. A. Malyshev and J. Knoester,
 Phys. Rev. Lett. {\bf 98}, 087401 (2007).

\bibitem{Klugkist2007} J. A. Klugkist, V. A. Malyshev, and J. Knoester,
     in preparation.


\end{thebibliography}
\end{document}